\documentclass[lettersize,journal]{IEEEtran}
\usepackage{amsmath,amsfonts}
\usepackage{algorithmic}
\usepackage{algorithm}
\usepackage{array}
\usepackage{textcomp}
\usepackage{stfloats}
\usepackage{url}
\usepackage{verbatim}
\usepackage{graphicx}
\usepackage{multirow}
\usepackage{makecell}
\usepackage{caption}
\usepackage{subfigure}
\usepackage{cite}
\usepackage{amsthm} % proof 所需包
\usepackage{amssymb}
\usepackage{color}
\usepackage{tabularx}
\usepackage{booktabs}
\newtheorem{lemma}{Lemma}

\newcolumntype{C}{>{\centering\arraybackslash}X}

\begin{document}

\title{Learning-Based Blockage-Resilient Beam Training in Near-Field Terahertz Communications}

\author{Caihao Weng, Yuqing Guo, Bowen Zhao, Ying Wang,~\IEEEmembership{Member,~IEEE},\\ Wen Chen,~\IEEEmembership{Senior Member,~IEEE,} and Zhendong Li 
\vspace{-0.7cm}
\thanks{This work was supported in part by the National Natural Science
Foundation of China under Grant 62531015 and Grant 62571064 and in part
by the BUPT Excellent Ph.D. Students Foundation under Grant CX20251019.}

\thanks{Caihao Weng, Yuqing Guo, and Ying Wang are with the State Key Laboratory
of Networking and Switching Technology, Beijing University of
Posts and Telecommunications, Beijing 100876, China (e-mail: wengcaihao@bupt.edu.cn; guoyuqing2020@bupt.edu.cn; wangying@bupt.edu.cn).}% <-this % stops a space

\thanks{Bowen Zhao is with National Engineering Research Center for Mobile Network Technologies, Beijing University of Posts and Telecommunications, Beijing 100876, China (e-mail: zhaobowen0421@bupt.edu.cn).}

\thanks{Wen Chen is with the Department of Electronic Engineering, Shanghai Jiao Tong University, Shanghai 200240, China (e-mail: wenchen@sjtu.edu.cn.)}

\thanks{Zhendong Li is with the School of Information and Communication Engineering, Xi'an Jiaotong University, Xi'an 710049, China (e-mail: lizhendong@xjtu.edu.cn).}
}

\markboth{Journal of \LaTeX\ Class Files,~Vol.~14, No.~8, August~2021}%
{Shell \MakeLowercase{\textit{et al.}}: A Sample Article Using IEEEtran.cls for IEEE Journals}


\maketitle

\begin{abstract}
Terahertz (THz) band is considered a promising candidate to meet the high-throughput requirement for future sixth-generation (6G) wireless communications due to its ultra-wide bandwidth. However, due to the high penetration loss at high-frequencies, blockage becomes a serious problem in THz communications, especially in near-field indoor communications with numerous obstacles. To address this issue, this paper investigates blockage-resilient near-field beam training based on self-accelerating Airy beam, which can propagate along a curved trajectory to circumvent obstacles. Specifically, we first analyze the trajectory of the Airy beam and the beam pattern at the receiver using a discrete Fourier transform (DFT) codebook in the presence of obstacles. Interestingly, we reveal that the beam pattern not only captures the receiver's location information but also implicitly encodes the spatial relationship between the receiver and obstacle, which facilitates identifying the optimal Airy beam configuration. Based on this insight, we formulate the blockage-resilient beam training task as a multi-task learning problem and propose a lightweight attention-based multi-parameter beam training network (AMPBT-Net) to jointly predict the angle, distance, and curvature parameters of the optimal Airy beam based on the beam pattern. Finally, simulation results demonstrate that the Airy beam effectively mitigates blockage effects and the proposed scheme achieves comparable performance to exhaustive beam sweeping while significantly reducing training overhead.
\end{abstract}


\section{Introduction}
The inexorable demand for higher data throughput and ultra-low latency in sixth-generation (6G) wireless communications has driven a shift toward the high-frequency band \cite{THz-1,THz-2}. In particular, the terahertz (THz) band, spanning from 0.1 to 10 THz, has emerged as a particularly promising frontier, offering vast swaths of underutilized bandwidth capable of supporting the terabit-per-second (Tbps) data rates envisioned for 6G and beyond \cite{hop-1, THz-3, THz-4}. 

This migration to higher frequencies, however, introduces profound implications for future 6G system design. In the THz band, the short wavelength permits packing a substantially larger number of antenna elements into a fixed physical aperture \cite{bending_1,sense_then_train}. This capability has driven the development of novel antenna architectures, such as extremely large-scale multiple-input-multiple-output (XL-MIMO) \cite{wch-tvt,wch-twc} and holographic MIMO (HMIMO)\cite{HMIMO_1,HMIMO_2}, which promise significant improvements in spatial multiplexing and beamforming precision.

More fundamentally, the move to higher frequencies enlarges the Rayleigh distance, which delineates the boundary between the near-field and far-field regions\cite{Rayleigh}. For instance, a modest 30 cm array aperture operating at 100 GHz creates a near-field region that extends up to 60 meters. This expansion makes near-field communication an unavoidable and defining characteristic of future 6G systems, and has therefore led to intensive research on near-field communications in recent years. In August 2025, 3rd Generation Partnership Project (3GPP) technical specification group (TSG) radio access
network (RAN) WG1 \#122 met in Bangalore, India. During the meeting, more than a dozen companies, among them Huawei and NVIDIA, submitted proposals addressing near-field issues, such as channel modeling and beam management, for consideration on the 6G agenda, underscoring that near-field communication has progressed beyond theoretical exploration to emerge as an indispensable practical challenge that must be systematically addressed in the evolution of 6G systems.\cite{3gpp-122}.

A salient feature of the emerging near-field paradigm is the diversity of available wavefronts\cite{wavefront-1,hop-2}. In particular, spherical wavefronts, which offer high beam gain, have attracted considerable attention. Unlike far-field planar wavefronts, spherical wavefronts can focus electromagnetic energy on a specific spatial point, rather than merely steering it toward a certain direction\cite{Tcom-DLL}. This additional degree of freedom (DoF) enables a substantial improvement in communication performance, albeit at the cost of increased complexity, especially in beam training and beam management \cite{wch-twc}. Thus, extensive efforts have been devoted to developing low-complexity solutions that balance performance and overhead in near-field communications.

\subsection{Related Works}
Two-stage search has been one of the most widely adopted methodologies for reducing beam training overhead in the near-field \cite{wch-tvt,hierarchical_1}. In this framework, the receiver angle is first determined using a discrete Fourier transform (DFT) codebook, after which a customized near-field codebook is employed to estimate the receiver distance. Then, two neural networks have been utilized in \cite{learning-1} to estimate the receiver angle and distance based on the received signals corresponding to the far-field DFT codebook. Subsequently, some studies have revealed that near-field beam training can still be realized solely with the far-field DFT codebook \cite{dft_bt_1,dft_bt_2}, even without the assistance of neural networks, leading to several unexpected yet intriguing findings. The core idea behind these methods is to analyze the beam pattern of the far-field DFT codebook and derive closed-form or semi-closed-form relationships between its characteristics (e.g., width or central beam gain) and the receiver's angle and distance, thereby enabling near-field beam training. In \cite{learning-1}, the relation has instead been learned  implicitly by neural networks in a black-box manner. These approaches in \cite{dft_bt_1,dft_bt_2} have been shown to not only substantially reduce training overhead but also provide off-grid distance estimation, thereby further improving communication performance. To address the practical hybrid far-field and near-field scenarios, a wavenumber-domain beam training method has been proposed in \cite{wch-twc,gyq_training}, which is based on decomposing any spherical wave into a superposition of plane waves. This method unifies beam training for both far- and near-field regions with low training overhead and yields substantial gain improvements, owing to its capability of utilizing arbitrary spherical waves instead of the limited ones used in conventional schemes.

However, a critical challenge often overlooked in these beam training frameworks is the profound vulnerability of THz links to blockage. Due to the inherently high penetration losses that escalate with frequency \cite{GlobeCOM-NN,NC-princeton}, even minor obstructions can severely attenuate or completely disrupt a communication link. This issue is particularly acute in indoor near-field environments, such as wireless data centers\cite{airy_codebook} and Internet of Things communications\cite{wavefront-1}. Although several techniques have been proposed to mitigate blockage, such as pinching antennas\cite{pinching_1,pinching_2}, movable antennas\cite{movable_1,movable_2}, and reconfigurable intelligent surfaces (RISs)\cite{RIS_1,RIS_2,RIS_3}, their core strategy relies on physically altering the communication geometry, either by moving antenna elements or by introducing new reflective nodes to establish an alternative, unobstructed line-of-sight (LoS) path. The effectiveness of these methods, however, is contingent upon the availability of such auxiliary hardware, which may not be present in all deployment scenarios.

Fortunately, while this susceptibility to blockage originates from high-frequency operation, the resulting near-field region itself offers potential remedies through the exploitation of underexplored near-field wavefronts that exhibit distinctive propagation characteristics. Specifically, the Bessel beam demonstrates remarkable resilience against blockages, a capability directly attributed to its intrinsic properties of self-healing and non-diffracting properties \cite{wavefront-1}. Alternatively, the Airy beam, endowed with both self-healing and self-accelerating characteristics, can propagate along a curved trajectory to circumvent obstacles, a feature that has recently attracted some attention in near-field wireless communications. In \cite{NC-princeton,GlobeCOM-NN,commun_engineer}, both experimental and simulation results have demonstrated that, by appropriately configuring the parameters of the Airy beam, channel capacity can be significantly enhanced while blind spots in the coverage area are effectively reduced. Moreover, the Airy codebook has been first proposed in \cite{airy_codebook} and several low-complexity beam search schemes have been provided to establish the communication link between the transceiver. In addition, inspired by the curved trajectory of the Airy beam, the concept of wavefront hopping has been proposed to enhance near-field physical-layer security and mitigate inter-cell interference by dynamically switching beams along different trajectories over time \cite{hop-1,hop-2}.

\subsection{Motivations and Contributions}
Inspired by the high penetration loss of obstacles in high-frequency bands and the curved propagation trajectory of Airy beams, this paper investigates the important problem of near-field blockage-resilient beam training. Analogous to how a focusing beam achieves spatial focusing by incorporating a distance parameter into a steering beam, Airy beams can navigate around obstacles along a curved trajectory through the introduction of a curvature parameter\cite{GlobeCOM-NN, airy_codebook}. In the context of beam training, these additional parameters increase the codebook dimensionality and training overhead, a persistent problem in beam training. To address this issue, in this paper, a deep learning-based blockage-resilient beam training method is proposed. Specifically, our contributions are summarized as follows.

\begin{itemize}
\item[$\bullet$] First, Airy beam is introduced into near-field beam training to mitigate the impact of obstacles between the transceiver in high-frequency bands. We analyze the propagation trajectory of the Airy beam and, using the stationary phase approximation, derive both its trajectory representation under given parameters and a closed-form expression in the paraxial regime. Then, we investigate the beam pattern at the receiver obtained from the DFT codebook in the presence of obstacles. Interestingly, we reveal that the received beam pattern not only reflects the receiver’s angle and distance but also implicitly encodes the relationship between the receiver and obstacle positions, which are essential for determining the optimal Airy beam. Nevertheless, it is intractable to obtain an explicit analytical expression for this relationship. Thus, we propose a deep learning-based framework to infer the optimal Airy beam parameters directly from the beam patterns.

\item[$\bullet$] Second, the relationship between the beam pattern and the different parameters of the optimal Airy beam is modeled as a multi-task learning problem, where learning each parameter is treated as an independent task. To reduce the complexity of the proposed method, we propose an attention-based multi-parameter beam training network (AMPBT-Net). A lightweight attention mechanism is incorporated into AMPBT-Net to allow each task to selectively extract the most relevant features from the shared beam pattern representation, facilitating accurate estimation of the optimal Airy beam parameters while mitigating interference between tasks.

\item[$\bullet$] Finally, extensive simulations and performance evaluations are provided. Numerical results demonstrate that Airy beams can effectively mitigate the impact of obstacles on transceiver communications compared to traditional Focus beams. Moreover, numerical results are presented to demonstrate the proposed deep learning-based beam training method achieves performance nearly equivalent to exhaustive beam sweeping while substantially reducing training overhead, outperforming other baseline solutions.
\end{itemize}

\subsection{Organization and Notation}
\emph{Organization}: The rest of this paper is organized as follows. Section \uppercase\expandafter{\romannumeral2} first introduces the near-field wavefront engineering and revisits the electromagnetic wave propagation with blockage. Then, we analyze the trajectory and beam pattern of the Airy beam in Section \uppercase\expandafter{\romannumeral3} and propose the AMPBT-Net to estimate the parameters for the optimal Airy beam in Section \uppercase\expandafter{\romannumeral4}. Numerical results and conclusions are provided in Section \uppercase\expandafter{\romannumeral5} and Section \uppercase\expandafter{\romannumeral6}.

\emph{Notations}: Lower-case, bold lower-case, and bold upper-case letters correspond to scalars, vectors, and matrices, respectively. The complex numbers are denoted by $\mathbb{C}$. A complex Gaussian distribution with mean $\mu$ and covariance $\Sigma$ is written as $\mathcal{CN}(\mu, \Sigma)$. The notation $|\cdot|$ refers to the absolute value of a scalar or the cardinality of a vector. Finally, $\mathcal{F}\{\cdot\}$ and $\mathcal{F}^{-1}\{\cdot\}$ denote the Fourier and inverse Fourier transform operators, respectively.

\section{System Model}
In this section, we first introduce the considered near-field communication scenario. Next, we provide a brief review of three representative wavefronts in far-field and near-field regions, as well as electromagnetic wave propagation in the presence of blockages, to make the paper self-contained. Finally, the problem formulation is presented.

\subsection{Scenario}
We consider a THz narrowband near-field communication system in this paper. The transmitter is equipped with an $N$-element uniform linear array (ULA) and the receiver has a single antenna. Without loss of generality, the ULA is assumed to be located on the $y$-axis and be symmetric about the $x$-axis. Thus, the Cartesian coordinates of the $n$-th antenna is given by $\mathbf{p}_{t,n}=\left[0,y_{t,n}\right]^T=\left[0,n\delta\right]^T$, where $n\in\left\{\frac{1-N}{2},\ldots,0,\ldots,\frac{N-1}{2} \right\}$ and $\delta$ is the half-wavelength antenna spacing. The aperture of the ULA is represented by $L=(N-1)\delta$ and the Cartesian coordinates of the receiver are denoted by $\mathbf{p}_{r}=[x_{r},y_{r}]^T$.

\subsection{Wavefront Engineering From Far-field to Near-field}
In this subsection, three types of beams are introduced, namely the steering beam, focusing beam and Airy beam. Following the framework in \cite{GlobeCOM-NN}, the generation of these beams is characterized by a hierarchical parameter expansion based on a nested structure that incorporates the steering angle $\theta$, the focusing distance $r$, and the curvature coefficient $c$. Specifically, the phase required to generate the steering beam for the $n$-th transmit antenna is given by
\begin{equation}
\label{eq1}
\phi_{\text{steer}}(n,\theta) = -\kappa n \delta \sin \theta.
\end{equation}
where $\kappa=2\pi/\lambda$ denotes the wavenumber and $\lambda$ is the wavelength. Subsequently, an additional distance-dependent phase term is introduced into \eqref{eq1} to obtain the phase expression of the focusing beam as
\begin{equation}
\label{eq2}
\phi_{\text{focus}}(n,\theta,r) = \phi_{\text{steer}}(n,\theta) + \kappa \frac{\cos^2 \theta}{2r}n^2 \delta^2,
\end{equation}
where the expression is derived from the second-order truncated Taylor expansion of the Green’s function, whereas \eqref{eq1} represents the first-order truncated approximation\cite{wch-twc,cyb_wavenumber}.

Similarly, incorporating a curvature-dependent phase term into \eqref{eq2} to capture the self-acceleration property yields the phase expression of the Airy beam as
\begin{equation}
\label{eq3}
\phi_{\text{curve}}(n,\theta,r,c) = \phi_{\text{focus}}(n,\theta,r) - (2\pi c)^3 n^3 \delta^3 / 3,
\end{equation}
which can be synthesized in the Fourier plane by imposing a cubic phase at the transmitter\cite{airy_1}. Thus, the beam profile at the source aperture is expressed as
\begin{equation}
\label{eq4}
E_0(0,n\delta) = \frac{1}{\sqrt N}\exp(j\phi(n,\theta,r,c)).
\end{equation}

Unless otherwise specified, the initial phase $\phi$ in this paper refers to $\phi_\text{curve}$, which reduces to $\phi_{\text{focus}}$ when $c = 0$, and further degenerates to $\phi_{\text{steer}}$ when $c = 0$ and $r \to\infty$. This hierarchical parameter structure enables the Airy beam to be regarded as a natural extension of the focusing beam, rather than an entirely new form, thereby allowing the Airy beam to inherit and generalize the focusing behavior while introducing controllable curvature.

\subsection{Electromagnetic Wave Propagation under Blockage}
The Rayleigh-Sommerfeld theory \cite{book_1} states that the electric field at any point $(x,y)\in\mathbb{R}^2$, $E(x,y)$, can be obtained by superposing the contributions from all individual point sources as
\begin{equation}
\label{eq5}
    \begin{aligned}
        E(x,y\,\vert \, E_0) = \int_\mathcal{S}  E_0(0,y_0) \frac{e^{-j\kappa r_{\text{u}}}x}{2\pi r_{\text{u}}^2}\left(j\kappa +\frac{1}{r_\text{u}}\right) dy_0,
    \end{aligned}
\end{equation}
which is derived from Maxwell's equations by using Green’s theorem. $r_u=\sqrt{(y-y_0)^2+x^2}$ is the distance between point $(x,y)$ and the source point $(0,y_0)$. $E_0(0,y_0)$ denotes the initial E-field distribution on the source aperture $\mathcal{S}$.

Notably, the integral in \eqref{eq5} can be interpreted as a convolution. Thus, it can be simplified using the angular spectrum method (ASM), which exploits Fourier transform principles to substantially reduce the computational complexity, yielding
\begin{equation}
\label{eq6}
    \mathcal{F}\left\{E(x,y\,\vert \, E_0) \right\}  =  \mathcal{F}\left\{E_0(0,y_0) \right\} 
     \times H(\kappa_y),
\end{equation}
where $H(\kappa_y)$ denotes the transfer function that characterizes the electromagnetic response generated by a point source \cite{NirvaWave}, and is expressed as
\begin{equation}
\label{eq7}
    \begin{aligned}
        H(\kappa_y)  = & \mathcal{F}\left\{\frac{1}{2\pi}\frac{\partial}{\partial_{x}}\left(\frac{e^{-j\kappa r}}{r}\right)\right\} \\
        = & \exp\left(-j \kappa x \sqrt{1-\lambda^2 \kappa_y^2} \right)
        ,
    \end{aligned}
\end{equation}
where $\kappa_y$ represents the wavenumber (or spatial frequency) along the $y$-axis. By applying the inverse Fourier transform to both sides of \eqref{eq6}, the electric field $E(x,y\,\vert \, E_0)$ can be expressed as
\begin{equation}
\label{eq8}
E(x,y\,\vert \, E_0) = \mathcal{F}^{-1}\left\{\mathcal{F}\left\{E_0(0,y_0) \right\}
     \times H(\kappa_y)\right\}.
\end{equation}

In free space, the Rayleigh-Sommerfeld theory provides an accurate characterization of the electric field. Furthermore, the ASM in \eqref{eq6} enables the computation of the electric field $E(x,:)$, either directly from the initial field at $x=0$ in a single step or iteratively from the previously computed field at $x=x-\delta_x$ \cite{NirvaWave}. However, in the presence of blockages, the continuity of electromagnetic wave propagation is disrupted, and the ASM in \eqref{eq6} must be applied iteratively to account for the discontinuities in the electric field introduced by the obstructions. Following \cite{NirvaWave,GlobeCOM-NN}, a binary matrix $B(x,y)$ is employed to characterize environment blockages. More specifically, $B(x,y) = 1$ indicates the absence of a blockage at position $(x,y)$, whereas $B(x,y)=\alpha$ represents signal attenuation due to a blocker located at $(x,y)$, where $\alpha\in[0,1)$ is the attenuation coefficient. Accordingly, the electric field at the $i$-th iteration is expressed as
\begin{equation}
\label{eq9}
E(x,y) = B(x,y) \mathcal{F}^{-1}\left\{\mathcal{F}\left\{E_0(x-\delta_x,y_0) \right\} \times H(\kappa_y)\right\}.
\end{equation}

In the implementation of the ASM in \eqref{eq9}, the propagation environment is discretized onto a uniform two-dimensional (2D) grid. The sampling intervals along the $x$- and $y$-axes, denoted by $\delta_x$ and $\delta_y$, must not exceed half of the wavelength to ensure numerical accuracy. Thus, $x=i_x\times\delta_x$ and $i_x$ denotes the grid index. 
Based on the iterative process in \eqref{eq9}, the electric field at any position within the environment can then be computed.

\subsection{Problem Formulation}
Based on the electric field in \eqref{eq9}, the beam gain at the receiver can be expressed as \cite{beam_gain}
\begin{equation}
\label{eq10}
g = \vert E(x_{r},y_{r}\, \vert \, E_{0}) \vert^2,
\end{equation}
where $E_{0}$ is the transmitted electric field. Drawing on prior studies, one widely adopted strategy for maximizing the beam gain in \eqref{eq10} is beam training, which aims to identify the optimal codeword that maximizes the achievable rate from a predefined codebook $\mathcal{B}$. Therefore, the codeword selection problem can be formulated as \cite{hierar_DLL}
\begin{equation}
    \label{eq11}
    E_0^{\star} = \mathop{\arg\max}\limits_{E_0 \in \mathcal{B}}\ g.
\end{equation}

In this paper, the codebook is designed as
\begin{equation}
    \label{eq12}
    \mathcal{B}=\left\{\mathbf{b}\left(\theta_{l_1},r_{l_2},c_{l_3}\right) \big| \theta_{l_1}\in \Theta,r_{l_2}\in\mathcal{R},c_{l_3}\in \mathcal{C}\right\},
\end{equation}
where $\mathbf{b}\left(\theta,r,c\right)$ denotes an Airy codeword, which is expressed as
\begin{equation}
    \label{eq13}
    \mathbf{b}\left(\theta,r,c\right)=\frac{1}{\sqrt{N}}\left[e^{j\phi(\frac{1-N}{2},\theta,r,c)} ,\ldots,e^{j\phi(\frac{N-1}{2},\theta,r,c)}\right]^T.
\end{equation}

Here, $\Theta$, $\mathcal{R}$, and $\mathcal{C}$ are the collections of the sampled angles, distances, and curvatures, respectively. Without loss of generality, we assume that these parameters are uniformly sampled\footnote {The proposed beam training methods can be readily extended to the Airy beam codebook proposed in \cite{airy_codebook}, where the sampling intervals of angle, distance, and curvature are determined to achieve a target correlation between two Airy beams.}. Hence, the sampled values are given by
\begin{equation}
\label{eq14}
\begin{cases}
\theta_{l_1}=\arcsin\left(\sin\theta_{\text{min}}+(l_1-1)\frac{\sin\theta_{\text{max}}-\sin\theta_{\text{min}}}{L_1-1}\right),\\
r_{l_2} = r_{\text{min}} + (l_2-1)\frac{r_{\text{max}}-r_{\text{min}}}{L_2-1}, \\
c_{l_3}=-c_{\text{max}}+(l_3-1)\frac{2c_{\text{max}}}{L_3-1},
\end{cases}
\end{equation}
where $l_i\in [1,L_i],i=1,2,3$. $L_1$, $L_2$, and $L_3$ denote the number of samples for $\theta$, $r$, and $c$, respectively.

\section{Trajectory and Beam Pattern Analysis}
To gain deeper insights into near-field beam training using the Airy beam, we begin by analyzing the trajectory of the Airy beam generated by the phase in \eqref{eq3}. Next, we examine the beam pattern received by the receiver during beam sweeping using a far-field DFT codebook in the presence of an obstacle between the transceiver. Interestingly, we show that the beam pattern not only reflects the receiver’s angle and distance but also implicitly encodes the relative angular and distance relationship between the receiver and the obstacle.

\subsection{Trajectory Analysis}
\begin{figure*}[htbp]
\centering
\subfigure[\label{trajectory_1}]{ \includegraphics[width=5.6cm]{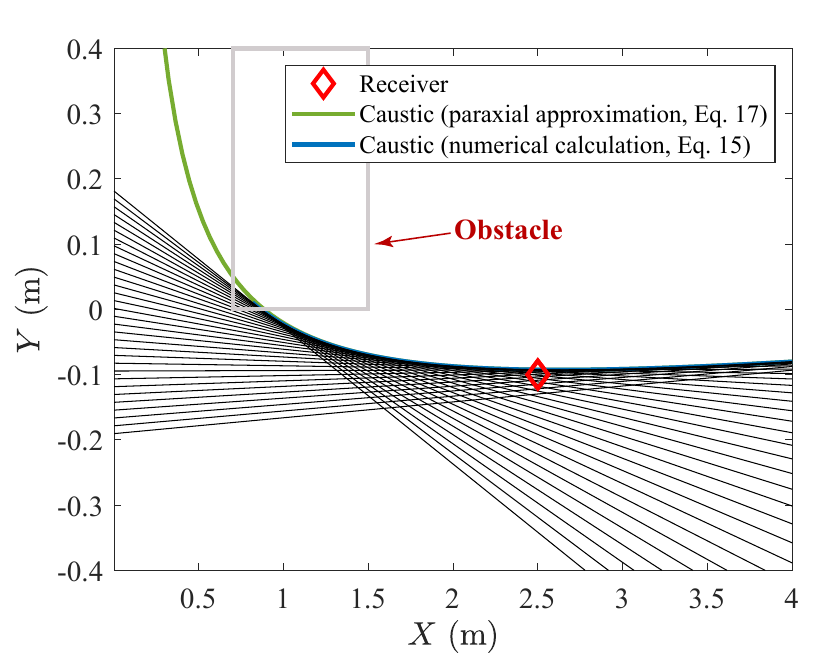}
}
\subfigure[\label{trajectory_2}]{
\includegraphics[width=5.6cm]{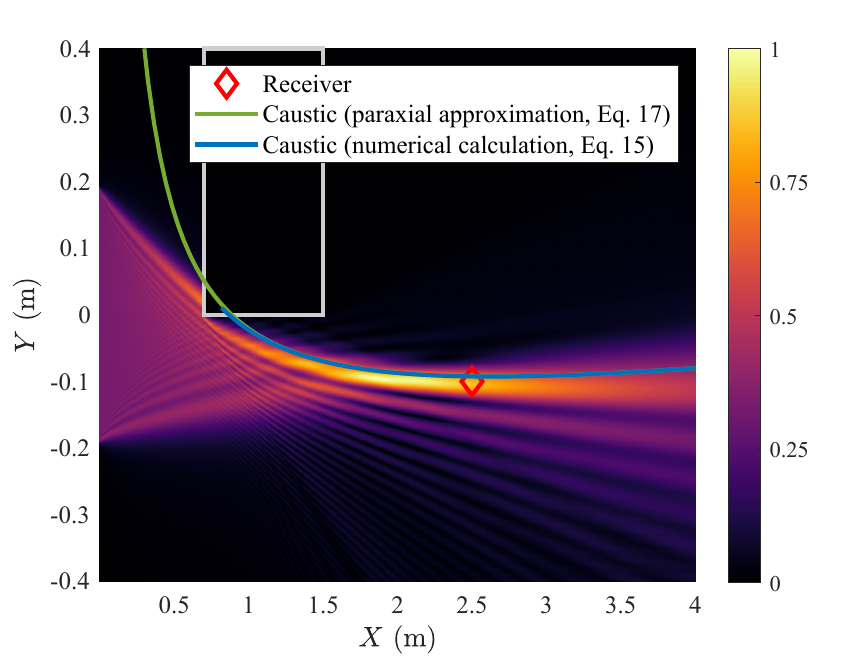}
}
\subfigure[\label{trajectory_3}]{
\includegraphics[width=5.6cm]{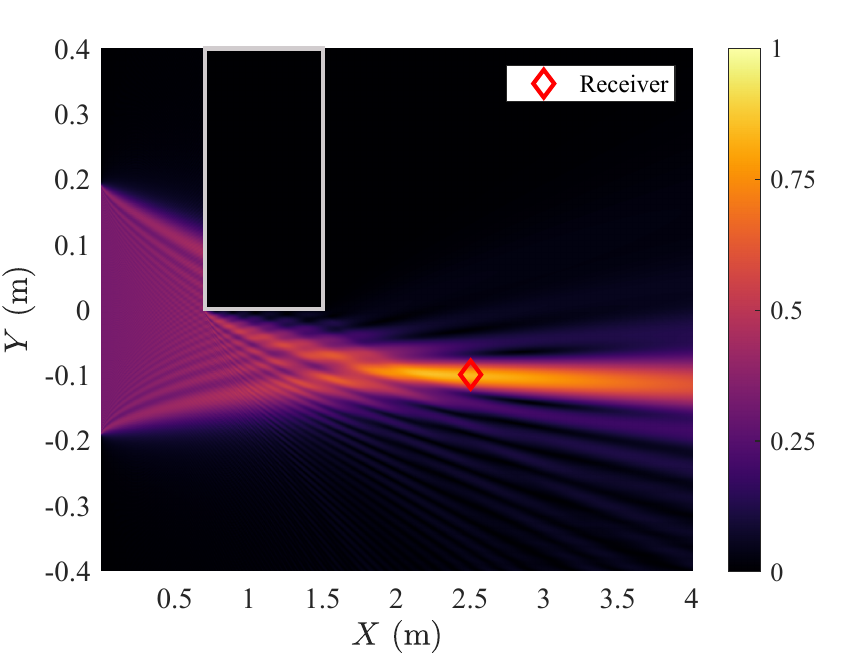}
}
\caption{Beamforming in the presence of blockage. (a) Ray representation. (b) Simulated propagation of the Airy beam. (c) Simulated propagation of the focusing beam. The carrier frequency is $f=100$ GHz, and the number of antennas is $N=255$.}
\label{compare}
\end{figure*}

In this paper, the beam trajectory is denoted as $y_c=f(x_c)$. While individual rays propagate strictly along straight paths and cannot bend, their collective envelopes (i.e., caustic) can form the desired curved trajectory, as illustrated in Fig. \ref{trajectory_1}. The caustic is the envelope formed by rays emerging from the antenna elements, with each ray tangent to the caustic. If the slope of the caustic at point $\left[x_c, y_c\right]$ is defined as $\tan \varphi$, then we have $df(x_c)/d_{x_c} = \tan \varphi = \sin\varphi / \sqrt{1-\sin^2\varphi}$.

Based on \eqref{app_eq2} in Appendix A, $\sin \varphi$ is rewritten as $\sin \varphi = \left(y_c - y_0 \right) / r_c = -\phi^{\prime}\left(y_0\right) / \kappa$, where $r_c=\sqrt{(y_c-y_0)^2+x_c^2}$ is the distance between point $[x_c,y_c]$ at the trajectory and the point $[0,y_0]$ at the source aperture. Thus, the essence of beam curving lies in shaping the desired trajectory by controlling the slope of the emitted ray from each antenna element through precise phase manipulation across the array. 

\begin{lemma}
If the transmitter employ the beam profile in \eqref{eq3}, the parametric expression of the trajectory $y_c=f(x_c)$ or caustic can be derived using the stationary phase approximation as
\begin{equation}
\label{eq15}
(x_c,y_c)= \left(\frac{\left(1-A(y_0)^2\right)^{3/2}}{B(y_0)},y_0-\frac{A(y_0)\left(1-A(y_0)^2\right)}{B(y_0)}\right),
\end{equation}
where $A(y_0)=\frac{1}{\kappa}\phi^{\prime}(y_0)=-\sin\theta+\frac{\cos^2\theta}{r_0}y_0-\frac{(2\pi c)^3 y_0^2}{\kappa}$, $B(y_0)=\frac{1}{\kappa}\phi^{\prime\prime}(y_0)=\frac{\cos^2\theta}{r_0}-\frac{2(2\pi c)^3 y_0}{\kappa}$, and $y_0$ denotes the point at the source aperture. 
\end{lemma}
\renewcommand\qedsymbol{$\blacksquare$}
\begin{proof}
The proof is provided in Appendix A.
\end{proof}
\renewcommand\qedsymbol{QED}

Due to the complexity of the initial phase in \eqref{eq3}, a closed-form analytical expression for $y_c=f(x_c)$ cannot be directly derived from \eqref{eq15}. Nevertheless, the caustic can be obtained and visualized using numerical calculation. Moreover, in the paraxial regime, where the angle between the wave propagation direction and the $x$-axis is very small, the trajectory can be approximated as stated in the following lemma \cite{stationary_phase_approximation}.

\begin{lemma}In the paraxial regime, where the propagation angle relative to the $x$-axis is small, $\sin\phi$ can be approximated by $\tan\phi$, and the parametric expression in \eqref{eq15} can thus be approximated as 
\begin{equation}
\label{eq16}
(x_c,y_c)= \left(\frac{\kappa}{\phi^{\prime\prime}(y_0)},y_0-\frac{\phi^{\prime}(y_0)}{\phi^{\prime\prime}(y_0)}\right),
\end{equation}
and the approximated closed-form trajectory in this paraxial regime is given by 
\begin{equation}
\label{eq17}
y_c=x_c\sin\theta - \frac{\kappa x_c\cos^4\theta}{32\pi^3c^3r^2} - \frac{\kappa}{32\pi^3c^3x_c} + \frac{\kappa \cos^2\theta}{16 \pi^3 c^3 r} .
\end{equation}
\end{lemma}
\renewcommand\qedsymbol{$\blacksquare$}
\begin{proof}
The proof is provided in Appendix B.
\end{proof}
\renewcommand\qedsymbol{QED}

Fig. \ref{trajectory_2} illustrates an example comparing the beam trajectory obtained from \eqref{eq15} (blue curve) and its paraxial approximation in \eqref{eq17} (green curve). Both curves closely follow the desired curved trajectory, demonstrating the accuracy of the analytical and approximate expressions. An Airy beam is generated with parameters $\theta=-0.047$, $r=1.589$ and $c=-2.246$ to maximize the power delivered to the receiver at $\mathbf{p}_{r}=[2.5, -0.1]^T$. A rectangular obstacle is placed at $[x_{\text{b}},y_{\text{b}}]=[1.2,0.2]$ with dimensions $(d_{\text{b},x},d_{\text{b},y})=(0.8,0.4)$. Fig. \ref{trajectory_3} repeats the same example using a focusing beam, the optimal parameters are $\theta=-0.035$ and $r=2.988$. Experiments in this example demonstrate that Airy beams can outperform classical the focusing beam by carefully designing the beam's trajectory via precise phase manipulation across the array, thereby navigating around obstacles. Moreover, the optimal Airy beam and the optimal focusing beam do not necessarily share the same angle and distance parameters.

Because the caustic is generated by rays emitted from the antenna elements, the maximum distance over which an Airy beam can maintain its prescribed trajectory is limited by the array aperture size $L$, specifically by the positions of the array edges. For the trajectory in \eqref{eq17}, a ray tangent to the trajectory at point $(x_c,y_c)$ is described analytically by 
\begin{equation}
\label{eq18}
\frac{y-y_c}{x-x_c}=\frac{dy_c}{dx_c} = \sin\theta - \frac{\kappa \cos^4\theta}{32\pi^3c^3r^2} + \frac{\kappa}{32 \pi^3 c^3 x_c^2}.
\end{equation}
The maximum distance over which an Airy beam can follow its prescribed trajectory, $x_{\text{max}}$, is determined by the rays originating from the edges of the aperture, located at $[0,\pm L/2]$. Denoting these edge points as $[x,y] = [0,\pm L/2]$ and expressing $y_c$ in \eqref{eq18} in terms of $x_c$ according to \eqref{eq17}, the maximum distance $x_{\text{max}}$ is expressed as
\begin{equation}
\label{eq19}
x_{\text{max}} = \max \left(\frac{\kappa r}{\kappa \cos^2\theta + 8 \pi^3 c^3 r L},\frac{\kappa r}{\kappa \cos^2\theta - 8 \pi^3 c^3 r L}\right),
\end{equation}
where the first term, with the plus sign in the denominator, corresponds to the ray from $[x,y] = [0, -L/2]$. Obviously, which term attains the maximum in \eqref{eq19} is determined by the sign of the curvature $c$. In Fig. \ref{trajectory_2}, the curvature $c$ is negative, hence $x_{\text{max}}=\frac{\kappa r}{\kappa \cos^2\theta + 8 \pi^3 c^3 r L}\approx 8.5$ meters.

\subsection{Pattern Analysis}
In this subsection, the normalized received signal strength at the receiver is characterized when a DFT codebook is employed for beam sweeping, where the received signal strength is defined as $\mathbf{y}_{\text{DFT}}=\vert E(x_r,y_r\vert E_0) \vert$ and $E_0=1/\sqrt{N}\exp\left(j\phi_{\text{steer}}\right)$ here. 
\begin{figure}[htbp]
    \centering
    \includegraphics[width=8cm]{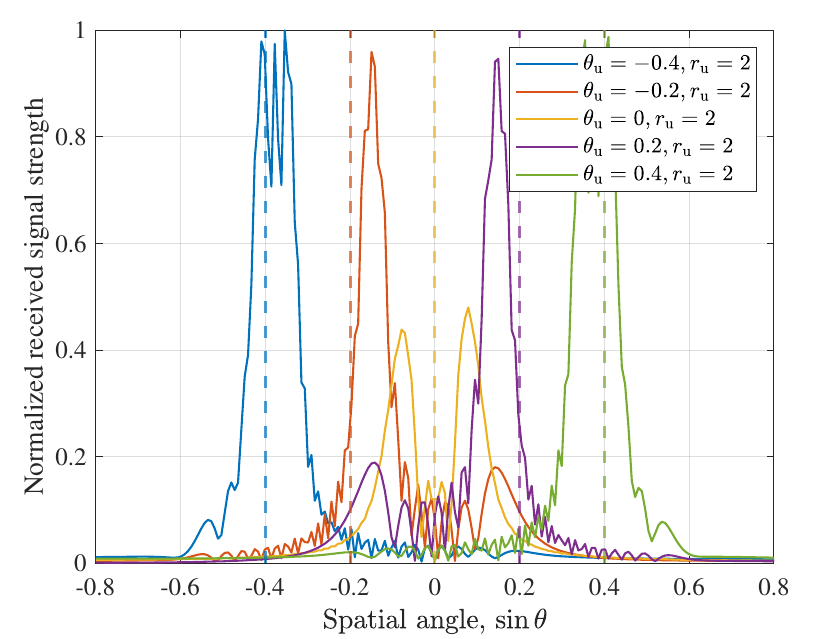}
    \caption{Normalized received signal strength for receivers at different spatial angles under DFT codewords, where $N=255$ and $f=100$ GHz.}  
    \label{pattern1}
\end{figure}
In Fig. \ref{pattern1}, we first present the normalized received signal strength under DFT codewords versus the spatial angle $\sin\theta$. There is a rectangular obstacle centered at $[x_{\text{b}},y_{\text{b}}]=[0.5,0]^T$, which corresponds to a spatial angle of zero, and has dimensions $(d_{\text{b},x},d_{\text{b},y})=(0.2,0.2)$. The dashed lines in the figure indicate the spatial angles of the corresponding receivers. Several key observations can be drawn as follows.

Existing studies have shown that, in the absence of obstacles, a receiver’s spatial angle is approximately located at the center of the dominant-angle region \cite{fast_bt,dft_bt_1}, which is defined as the range of spatial angles where the received signal strength obtained using the DFT codebook exceeds a predefined threshold. This straightforward relationship allows the center angles of the dominant-angle region to be used as candidate spatial angles for the optimal focusing beam.

However, as shown in Fig. \ref{pattern1}, the presence of an obstacle disrupts this behavior. The receiver’s spatial angle is no longer necessarily located at the center of the dominant-angle region. Instead, the mapping between the receiver's angle and the received signal becomes more complex (i.e., Eq. \ref{eq9}). Furthermore, Fig. \ref{pattern1} shows that as the receiver’s spatial angle approaches that of the obstacle, i.e., as the receiver becomes increasingly occluded, the conventional center-based heuristic gradually breaks down. In addition, due to the high penetration loss in high-frequency bands \cite{THz-5}, the received signals observed at spatial angles blocked by an obstacle experience significant attenuation. This phenomenon is referred to as beam pattern collapse in this paper. 

\begin{figure}[htbp]
    \centering
    \includegraphics[width=8cm]{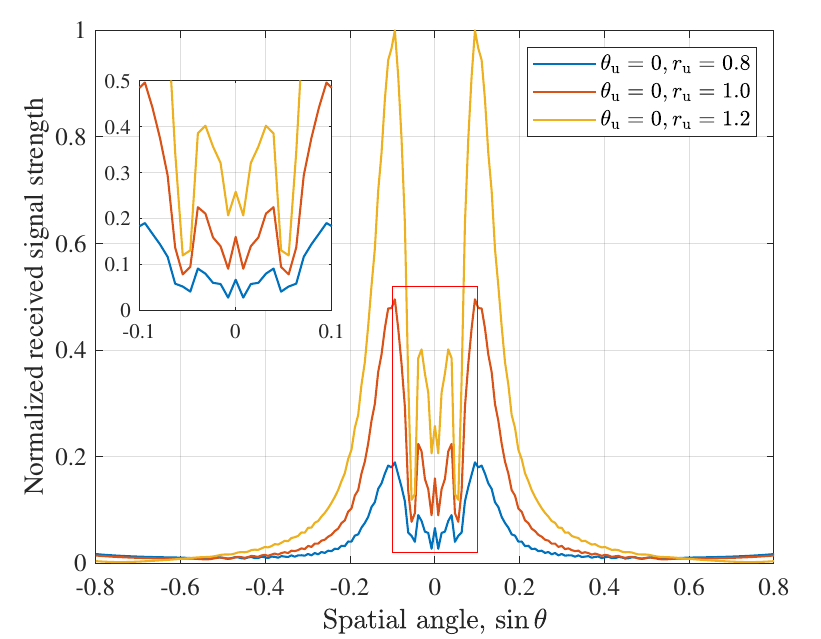}
    \caption{Normalized received signal strength for receivers at different distances under DFT codewords, where $N=255$ and $f=100$ GHz.}  
    \label{pattern2}
\end{figure}

Fig. \ref{pattern2} shows the normalized received signal strength versus spatial angle for receivers at different distances, under the same obstacle conditions as in Fig. \ref{pattern1}. It can be observed that when the LoS path between the transmitter and the receiver is occluded by an obstacle, receivers at different distances, or equivalently, at different distances from the obstacle, exhibit distinct beam pattern.

In summary, based on the above analysis and \eqref{eq9}, when a receiver is blocked by an obstacle, the DFT beam pattern captures the coupled influence of both the receiver and the obstacle. Specifically, the beam pattern not only reflects the receiver’s spatial angle and distance but also implicitly encodes the relative angular and distance relationship between the receiver and the obstacle, providing informative cues that can be exploited to identify the optimal beam for beam training. Nevertheless, it is intractable to derive an explicit analytical expression for this relationship. Motivated by this, in the following section, we propose a deep learning-based beam training scheme designed to capture the relationship between the DFT beam pattern and the parameters of the optimal Airy beam.

\section{The proposed AMPBT-Net}
In this section, we propose a deep learning-based beam training scheme designed to estimate the parameters of the optimal Airy beam. We first present the motivation for incorporating the attention mechanism, followed by a detailed description of the architecture and design of the proposed AMPBT-Net. Finally, the deep learning-based beam training procedure is presented.

\subsection{Motivation}
A straightforward solution to predicting the parameters of the optimal Airy beam (i.e., $\theta$, $r$, and $c$) is to train three independent single parameter beam training networks (SPBT-Nets). In this scheme, each network takes the DFT beam pattern as input and is dedicated to predicting one specific parameter. However, treating each parameter prediction as an isolated task introduces several inherent limitations. First, it neglects the intrinsic correlations among the parameters, leading to redundant feature extraction across different networks. This not only increases the number of model parameters and computational overhead but also complicates the training and deployment processes. Second, the absence of cross-task information sharing and mutual constraints prevents each network from learning a more generalizable shared representation. Consequently, the individual networks are prone to overfitting under complex problems, resulting in suboptimal prediction accuracy.

To overcome these limitations, we reformulate the joint prediction of the optimal Airy beam parameters as a multi-task learning (MTL) problem, enabling efficient prediction of all parameters through a single unified network. Inspired by the attention mechanism in \cite{MTAN}, we propose a lightweight AMPBT-Net. The core architecture of AMPBT-Net consists of a shared backbone network for constructing a global feature pool and multiple lightweight task-specific attention modules for individual parameter prediction tasks. These attention modules learn to generate soft attention masks that are applied to the shared features, dynamically filtering and refining the most relevant task-oriented features for each prediction task.

\subsection{Architecture of AMPBT-Net}
\begin{figure*}[htbp]
    \centering
    \includegraphics[width=17cm]{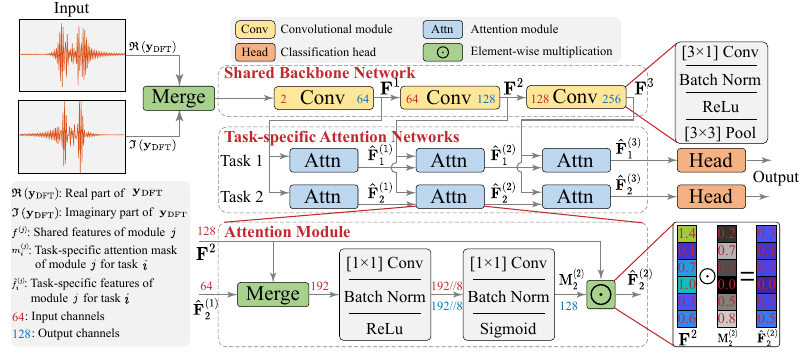}
    \caption{Visualization of the proposed AMPBT-Net. For clarity, the number of tasks $I$ is set to 2. The red and blue numbers in the figure denote the numbers of input and output channels of each module, respectively. The notation $[k\times k]$ represents a convolutional layer with a kernel size of $k\times k$.}  
    \label{network}
\end{figure*}

A detailed visualization of the proposed AMPBT-Net is shown in Fig. \ref{network}, which is composed of three principal components, a shared backbone network, $I$ parallel task-specific attention networks, and $I$ independent classification heads, where $I$ denotes the total number of tasks. 

The shared backbone network is implemented as a convolutional neural network (CNN) that serves as a universal feature extractor. It consists of $J=3$ sequential convolutional modules, each composed of a one-dimensional convolution (Conv1D) layer, a one-dimensional batch normalization (BatchNorm1D) layer, a ReLU activation function, and a one-dimensional max-pooling (Pool1D) layer. This backbone progressively transforms the raw beam pattern into higher-level shared feature maps, denoted as $\mathbf{F}^{j}$ for $j=1,\ldots,J$.

The task-specific attention networks contain $J$ attention modules that are structurally aligned with the $J$ convolutional modules in the shared backbone network. The primary function of these attention modules is to apply a soft attention mask $\mathbf{M}_i^{(j)}$ to the corresponding shared feature map $\mathbf{F}^{j}$, thereby generating task-specific features $\hat{\mathbf{F}}_i^{(j)}$ for task $i$ \cite{MTAN}, where $i=1,\ldots,I$ and $I=3$ for the Airy beam. Through this mechanism, each task adaptively emphasizes the most relevant portions of the shared features while suppressing irrelevant information.

Finally, the task-specific classification head is a fully connected (FC) layer that maps the refined task-specific features to the final output logits for each task.

\subsection{Detailed Design of AMPBT-Net}
In this subsection, the essential details of the proposed AMPBT-Net are first elaborated. Then, the complexity analysis is given. 

\subsubsection{AMPBT-Net Processing Flow}
Since the received signals under the DFT codebook are complex-valued, we first separate them into their real and imaginary parts, which are then concatenated and fed into the shared backbone network of AMPBT-Net for shared features extraction. Thus, the shared features $\mathbf{F}^{j}$ can be expressed as
\begin{equation}
\label{eq20}
\mathbf{F}^{j} = \text{Conv}^{(j)}\left(\mathbf{F}^{j-1}\right), j = 1,\ldots,J,
\end{equation}
where $\mathbf{F}^{0}=\left[\Re\left\{{\mathbf{y}_{\text{DFT}}}\right\};\Im\left\{{\mathbf{y}_{\text{DFT}}}\right\}\right]$ and $\text{Conv}^{(j)}$ denotes the operation of the $j$-th convolutional module. This process creates a global feature pool from which all parameter prediction tasks can extract relevant information.

As illustrated in Fig. \ref{network}, the shared features from the $j$-th convolutional module are subsequently fed into the corresponding attention module of each task-specific network to generate an attention mask $\mathbf{M}_i^{(j)}$, which is written as
\begin{equation}
\label{eq21}
\mathbf{M}_i^{(j)}=\text{Attn}_i^{(j)}\left( \left[\mathbf{F}^{j};\hat{\mathbf{F}}_i^{(j-1)}\right]\right),
\end{equation}
where $\text{Attn}_i^{(j)}$ is the operation of the $j$-th attention module for task $i$. Each attention module mainly consists of two lightweight sub-modules, which are both $[1 \times 1]$ kernel convolutional layers with batch normalization and a non-linear activation (ReLU or Sigmoid). It is worth noting that the input to attention module $j$ $(j>1)$ is not limited to the shared feature $\mathbf{F}^{j}$ from the current layer, but also incorporates the task-specific features $\hat{\mathbf{F}}_i^{(j-1)}$ from the previous attention module, which is expressed as
\begin{equation}
\label{eq22}
\hat{\mathbf{F}}_i^{(j)}=\mathbf{F}^{j} \odot \mathbf{M}_i^{(j)},i=1,\ldots,I,j=1,\ldots,J,
\end{equation}
where $\odot$ is the element-wise multiplication. This cascading design endows the feature selection process with contextual awareness, creating a progressive, layer-by-layer refinement stream for each task's unique features.

Finally, the output of the task-specific network is fed into the classification head to generate the probability distribution vector as
\begin{equation}
\label{eq23}
\hat{\mathbf{p}}_i^{b} = \text{Head}_i\left(\hat{\mathbf{F}}_i^{(3)}\right),
\end{equation}
where $\text{Head}_i$ is the operation of the classification head and $b$ is the index of samples in a batch.

\subsubsection{Loss Function} Since all task-specific heads in the proposed AMPBT-Net act as classifiers, a cross-entropy loss function is employed for each head, which is defined as
\begin{equation}
\label{eq24}
\mathcal{L}_i = -\frac{1}{B}\sum_{b=1}^{B}\sum_{l_i=1}^{L_i}p_{i,l_i}^{b}\log_{10}\hat{p}_{i,l_i}^{b},
\end{equation}
where $B$ and $L_i$ are the batch size and the total number of categories for task $i$, respectively. $p_{i,l_i}^{b}$ is a binary indicator that equals 1 if the $b$-th sample belongs to class $l_i$, and $0$ otherwise. $\hat{p}_{i,l_i}^{b}$ represents the $l_i$-th element of the predicted probability vector $\hat{\mathbf{p}}_i^{b}$, indicating the predicted probability that the $b$-th sample belongs to class $l_i$.

Therefore, the total multi-task loss is defined as 
\begin{equation}
\label{eq25}
\mathcal{L} = \sum_{i=1}^{I}\lambda_i\mathcal{L}_i,
\end{equation}
where $\lambda_i$ is a non-negative weighting coefficient that balances the contribution of each task. Two weighting strategies, namely equal weights (EW) and dynamic weight average (DWA) \cite{MTAN}, will be discussed in Section \uppercase\expandafter{\romannumeral5}.

\subsubsection{Complexity Analysis}
To evaluate the computational efficiency of the proposed AMPBT-Net, we perform a quantitative complexity analysis and compare it against a baseline of training $K$ independent SPBT-Nets. As the classification heads are identical in both configurations, our analysis focuses solely on the backbone and the attention network to provide a fair comparison of the core architectural efficiency. 

The complexity is measured in terms of the number of trainable parameters and multiply-accumulate operations (MACs). For the core components, the number of parameters in a Conv1D layer is calculated as $C_{\text{in}}\times C_{\text{out}}\times v + C_{\text{out}}$, and for a BatchNorm1D layer as $2\times C_{\text{out}}$. The computational cost of a Conv1D layer is approximately $C_{\text{in}}\times C_{\text{out}} \times v \times L_{\text{out}}$ MACs, where $C_{\text{in}},C_{\text{out}}$, $v$, and $L_{\text{out}}$ denote the number of input channels, the number of output channels, the convolution kernel size, and the length of the output feature map, respectively.

Based on the network configuration in Fig. \ref{network}, the shared backbone, which consists of three convolutional modules, contains 124,608 parameters and contributes around 0.005 Giga MACs (GMACs), assuming an input length of 255. The lightweight attention module, composed of efficient $[1\times1]$ convolutions, adds 41,008 parameters and 0.0018 GMACs per task-specific stream. For a comparative evaluation with $K=3$ tasks, the proposed AMPBT-Net aggregates to 247,632 parameters and 0.01 GMACs in total. In contrast, a benchmark consisting of three independent SPBT-Nets, each replicating the entire backbone, requires 373,824 parameters and 0.015 GMACs. 

Therefore, the AMPBT-Net achieves superior computational efficiency, requiring fewer parameters and MACs than the benchmark of training three independent networks. This dual advantage in both model size and computational cost demonstrates that AMPBT-Net is well-suited for efficient training and low-latency deployment in practical beam training systems.

\subsubsection{Algorithm Description}
To mitigate the prohibitive overhead associated with exhaustive sweeping of large Airy codebooks, we propose an efficient, learning-based beam training framework, which leverages a multi-task neural network to predict the optimal beam parameters. The entire process is divided into two distinct phases, an offline training phase and an online deployment phase.

In the offline training phase, the goal is to train the proposed AMPBT-Net to accurately learn the complex mapping from the DFT beam pattern $\mathbf{y}_{\text{DFT}}$ (network input) to the ground-truth indices of the optimal Airy beam codeword $\left(l_1^{\star},l_2^{\star},l_3^{\star}\right)$ (network output). 

The online deployment phase is designed for fast and efficient beam training in real-time scenarios. The procedure commences with the far-field beam training using the DFT codebook, producing the DFT beam pattern that serves as the network input. As described in Section \uppercase\expandafter{\romannumeral3}-C1, during inference, the trained AMPBT-Net outputs three probability vectors, $\hat{\mathbf{p}}_i^b,i=1,2,3$, each representing the likelihood distribution over all possible indices for its corresponding beam parameter. 

To improve robustness against potential prediction inaccuracies, the top $\hat{L}_i$ indices from each $\hat{\mathbf{p}}_i^b$ are selected to form the candidate sets, which are then combined to construct a compact candidate Airy beam codebook, $\mathcal{B}_{\text{candidate}}$. Finally, a rapid beam sweeping procedure is performed exclusively over this candidate codebook, enabling accurate identification of the optimal Airy beam codeword while dramatically reducing the beam training overhead from $L_1L_2L_3$ to $L_1+\hat{L}_1\hat{L}_2\hat{L}_3$.

\section{Simulation Results}
In this section, simulation results are presented to verify the resilience of the Airy beam against blockages and the effectiveness of the proposed AMPBT-Net. The transmitter is equipped with $N=255$ antennas, and the carrier frequency is set to $f=100$ GHz. The sampling intervals along both the $x$- and $y$-axes are $\delta_x=\delta_y=\lambda/2$, and the attenuation coefficient is $\alpha=0$.

The receiver moves within the region $x_r\in(0.5,4)$ m and $y_r\in(-2,2)$ m. Unless otherwise specified, a rectangular obstacle is centered at $[x_b,y_b]=[0.5,0]$ with dimensions $(d_{\text{b},x},d_{\text{b},y})=(0.2,0.2)$. For the Airy codebook, the number of samples for the angle, distance, and curvature parameters are $L_1=255$, $L_2=10$, and $L_3=51$, respectively. Specifically, the angle $\theta$ is uniformly sampled over $[-\pi/2,\pi/2]$, while the distance $r$ and curvature $c$ are uniformly sampled over $[0.5 ,5]$ and $[-5,5]$, respectively. For the proposed beam training method, the numbers of the candidate indices are set to $\hat{L}_1=3$, $\hat{L}_2=3$, and $\hat{L}_3=5$.

The dataset contains more than 40,000 samples, with $80\%$ used for training and $10\%$ each allocated to validation and testing. The batch size and learning rate are set to 64 and $2\times10^{-4}$, respectively.

For performance comparison, the following methods are considered.

1) \textbf{Airy-DL (proposed)}: The proposed deep learning-based beam training method employs the Airy codebook to mitigate the detrimental effects of obstacles on transceiver communication performance while maintaining low training overhead.

2) \textbf{Focus-DL}: 
The focusing codebook is used in the deep learning-based beam training method proposed in Section \uppercase\expandafter{\romannumeral4}.

3) \textbf{Airy-BS}: This benchmark performs exhaustive beam sweeping over the Airy codebook, representing the performance upper bound achievable with Airy beams.

4) \textbf{Focus-BS}: An exhaustive search using the near-field focusing codebook is conducted, serving as the performance upper bound for near-field focusing beams.

5) \textbf{Airy-Hier}: The hierarchical beam search strategy proposed in \cite{airy_codebook}, which first employs the focusing codebook to coarsely estimate the receiver’s angle and distance, and then refines the Airy beam’s curvature parameter in the second stage. This two-step process effectively reduces the training overhead from $L_1L_2L_3$ to $L_1L_2+L_3$.

Before conducting the overall performance comparison, several deep learning-based benchmarks `are introduced to evaluate the effectiveness of the proposed AMPBT-Net.

1) \textbf{SPBT-Nets}: This benchmark trains three completely separate and independent single-parameter beam training networks, one for each of the $I=3$ tasks, as described in Section \uppercase\expandafter{\romannumeral4}-A.

2) \textbf{EW}: In this scheme, the proposed AMPBT-Net is trained using equal task weights, i.e., $\lambda_i=1,\forall i$.

3) \textbf{DWA}: In this scheme, the proposed AMPBT-Net is trained using the adaptive weighting method proposed in \cite{MTAN}. Specifically, the weight $\lambda_i$ for task $i$ is computed as
\begin{equation}
\label{eq26}
\lambda_i(t) = \frac{I\exp\left(\omega_i(t-1)/T\right)}{\sum_i\exp\left(\omega_i(t-1)/T\right)},
\end{equation}
where $t$ and $T=2$ are the iteration index and a temperature which controls the softness of task weighting\cite{MTAN}. $\omega_i$ denotes the relative descending rate and is expressed as
\begin{equation}
\label{eq27}
\omega_i(t-1)=\frac{\mathcal{L}_i(t-1)}{\mathcal{L}_i(t-2)}, t \ge 3,
\end{equation}
and $\omega_i(1)=\omega_i(2)=1$.

\begin{table}[t]
\centering
\renewcommand\arraystretch{1.2}
\caption{Comparison of average beam gains of different deep learning methods. The optimal performance in each scenario is highlighted in bold.}
\label{learning_results}
\begin{tabular}{c|c|c|c|c}
\Xhline{1pt}
Beam type & \makecell{Obstacle \\ height (m)} & \makecell{AMPBT-Net, \\ EW} & \makecell{AMPBT-Net, \\ DWA} & SPBT-Nets \\ 
% 表头下的线也加粗
\Xhline{1pt}
\multirow{7}{*}{\makecell{Airy \\ beam}}      
 & 0.0 & 112.08 & $\textbf{112.14}$  &110.17 \\ \cline{2-5}
 & 0.1 & 96.07 & $\textbf{96.71}$ & 94.56 \\ \cline{2-5}
 & 0.2 & 85.75 & $\textbf{86.31}$ & 83.40 \\ \cline{2-5}
 & 0.3 & 77.35 & $\textbf{78.02}$ & 76.43 \\ \cline{2-5}
 & 0.4 & 73.23 & $\textbf{73.73}$ & 72.04 \\ \cline{2-5}
 & 0.5 & 67.18 & $\textbf{67.35}$ & 66.55 \\ \cline{2-5}
 & 0.6 & 62.65 & $\textbf{62.82}$ & 61.95 \\ \hline % 中间的分割线保持原样
\multirow{7}{*}{\makecell{Focusing \\ beam}}   
 & 0.0 & $\textbf{69.41}$ & 69.33 & 68.47 \\ \cline{2-5}
 & 0.1 & $\textbf{60.86}$ & 60.20 & 58.84 \\ \cline{2-5}
 & 0.2 & $\textbf{52.66}$ & 52.51 & 51.94 \\ \cline{2-5}
 & 0.3 & $\textbf{47.00}$ & 46.93 & 46.58 \\ \cline{2-5}
 & 0.4 & $\textbf{45.41}$ & 45.27 & 44.17 \\ \cline{2-5}
 & 0.5 & $\textbf{40.54}$ & 40.45 & 38.67 \\ \cline{2-5}
 & 0.6 & $\textbf{38.57}$ & 38.21 & 37.79 \\ 
% 使用 \Xhline{1pt} 来设置加粗的底线
\Xhline{1pt}
\end{tabular}
\end{table}

The comparison of the average beam gain achieved by the aforementioned deep learning methods is summarized in Table \ref{learning_results}. The results demonstrate that the proposed AMPBT-Net consistently outperforms the SPBT-Nets benchmark across all evaluated scenarios, which is a noteworthy finding. This indicates that, compared with SPBT-Nets, AMPBT-Net not only achieves lower computational complexity but also delivers superior prediction performance. Such improvement can be attributed to the cross-task information sharing and mutual constraints inherent in MTL, which enable the network to learn more generalized shared representations and mitigate overfitting in complex problems, which is a common advantage observed in MTL\cite{MTAN}. Furthermore, a comparison between the two weighting strategies, EW and DWA, shows that EW achieves better performance for conventional focusing beams. In contrast, for the Airy beams investigated in this study, DWA yields higher beam gain. This indicates that the learning dynamics of the curvature parameter differ from those of the angle and distance parameters, making the adaptive weighting mechanism of DWA more effective than the simple EW scheme.

\begin{figure}[htbp]
    \centering
    \includegraphics[width=8.7cm]{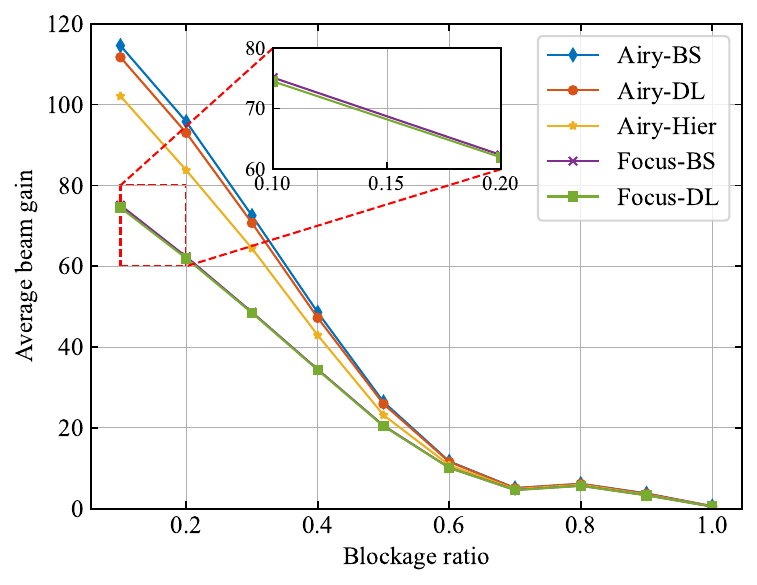}
    \caption{Average beam gain comparison with respect to the blockage ratio. Each point represents the average gain computed over a 0.05-width blockage ratio bin, plotted against the upper bound of the bin.}  
    \label{blockage}
\end{figure}

In Fig. \ref{blockage}, the impact of the blockage ratio on different beam training methods is illustrated, where the blockage ratio is defined as the proportion of the direct LoS path between a transmitting and a receiving antenna that is obstructed by obstacles. It is evident that Airy beam configurations achieve substantially higher average beam gains than their conventional focusing counterparts, particularly at low to moderate blockage ratios. This performance advantage gradually diminishes as the obstruction increases and becomes marginal when the blockage ratio exceeds 0.7. The proposed deep learning-based approaches (Airy-DL and Focus-DL) exhibit strong effectiveness, achieving near-optimal performance that closely approaches the upper bounds established by exhaustive beam sweeping. The slightly larger performance gap between Airy-DL and its upper bound, compared with that of the focusing beam, can be attributed to the increased complexity of learning the additional curvature parameter. Moreover, the proposed Airy-DL consistently outperforms the Airy-Hier in \cite{airy_codebook}, further validating its superiority.

\begin{figure}[htbp]
    \centering
    \includegraphics[width=8.7cm]{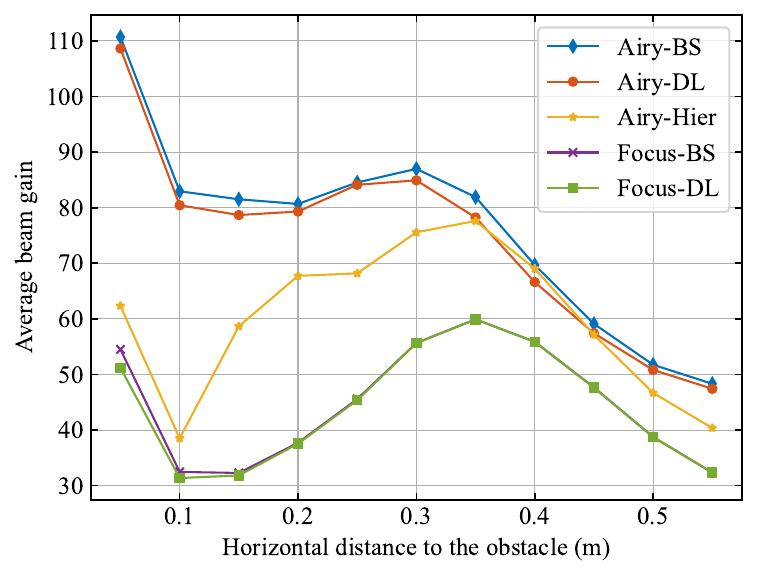}
    \caption{Average beam gain comparison with respect to the horizontal distance to the obstacle. Each point represents the average gain computed over a 0.05-width horizontal distance to the obstacle bin, plotted against the upper bound of the bin.}  
    \label{horizontal}
\end{figure}
Fig. \ref{horizontal} illustrates how the average beam gain varies with the horizontal distance between the receiver and the right edge of the obstacle under occlusion conditions, i.e., when the blockage ratio is greater than zero. It can first be observed that since the Airy beam can circumvent obstacles to transmit more energy to the receiver's location, it outperforms the focusing scheme at different horizontal distances. Moreover, the proposed learning-based method achieves performance close to that of exhaustive beam sweeping and, in most cases, surpasses Airy-Hier. This is because Airy-Hier performs an exhaustive curvature search based on the optimal focusing beam to derive the corresponding Airy beam. However, as presented in \uppercase\expandafter{\romannumeral3}-A, since the optimal angle and distance parameters of the focusing and Airy beams are not necessarily identical, Airy-Hier often fails to reach the true optimum even when all curvature values are exhaustively searched.

\begin{figure}[htbp]
    \centering
    \includegraphics[width=8.7cm]{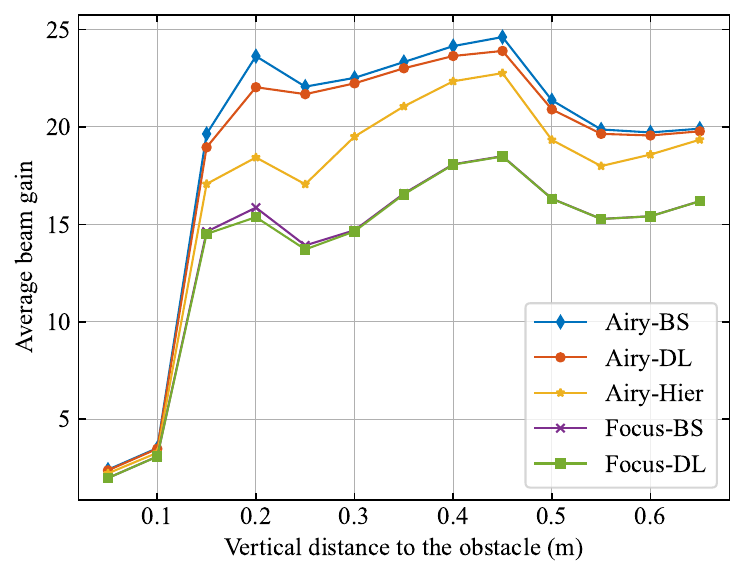}
    \caption{Average beam gain comparison with respect to the blockage ratio. Each point represents the average gain computed over a 0.05-width vertical distance to the obstacle bin, plotted against the upper bound of the bin.}  
    \label{vertical}
\end{figure}

In Fig. \ref{vertical}, we present the average beam gain performance as a function of the vertical distance between the receiver and the center of the obstacle, which increases from 0.05 m to 0.65 m. When the receiver’s vertical distance to the obstacle center is smaller than 0.10 (noting that the obstacle spans vertical coordinates from -0.1 to 0.1), the receiver lies almost directly behind the obstacle and thus experiences the most severe blockage. In this regime, the average beam gain is considerably lower than in other cases, and the differences among the evaluated methods are small. Once the vertical distance exceeds 0.10, a distinct transition occurs, the average beam gain of all methods increases sharply, and the Airy beam exhibits a clear performance advantage. These results indicate that while the Airy beam offers limited benefit under severe blockage, it provides significant improvement when the obstruction is moderate or relatively weak.

\begin{figure}[htbp]
    \centering
    \includegraphics[width=8.7cm]{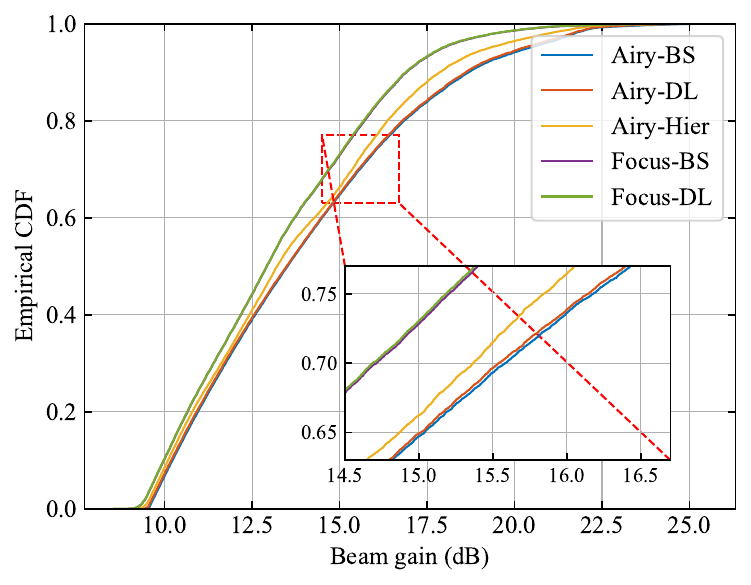}
    \caption{Empirical cumulative distribution function (CDF) of beam gain.}  
    \label{cdf}
\end{figure}
The overall beam gain performance of different methods under blockage conditions is illustrated in Fig. \ref{cdf}. The results clearly demonstrate the remarkable resilience of the Airy beam in maintaining high beam gain despite the presence of obstacles. At the median of the cumulative distribution function (CDF = 0.5), the Airy beam achieves a gain of about 13.5 dB, outperforming the focusing beam, which reaches 12.9 dB. This performance advantage becomes even more pronounced at higher beam gain levels, where receivers experience weaker blockage, consistent with the trends shown in Figs. \ref{blockage} and \ref{vertical}. Furthermore, the performance gap between exhaustive beam sweeping and the proposed deep learning variants within each beam type remains minimal, indicating that the learning-based approaches can effectively approximate the optimal beam selection.

\begin{figure}[htbp]
    \centering
    \includegraphics[width=8.7cm]{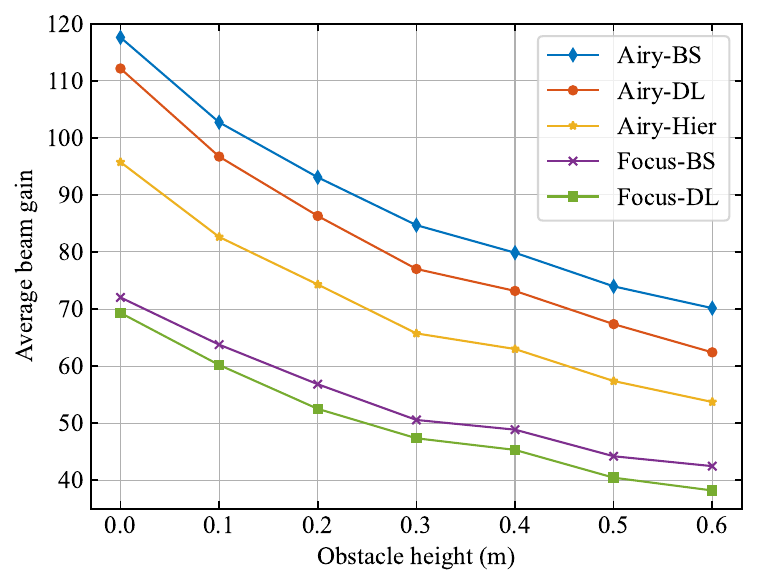}
    \caption{Average beam gain comparison with respect to the obstacle height.}  
    \label{height}
\end{figure}
In Fig. \ref{height}, the average beam gain performance of different methods against obstacle height is provided. The other simulation configurations are the same as those in Fig. \ref{blockage}. As evident from the figure, there is a monotonic decrease in the average beam gain for all methods with increasing obstacle height. This behavior is expected, as taller obstacles result in a greater proportion of receivers being obstructed and an increased blockage ratio. Notably, this downward trend gradually decelerates, and concurrently, the performance gap between the Airy and focusing beam narrows at greater obstacle heights. This phenomenon can be attributed to the saturation of the blockage effect. When an obstacle is sufficiently tall, especially given that the obstacle's center and the transmitting array's center are located on the same horizontal plane in our simulation, it can obstruct the entire transmitter aperture. Based on the principle of Airy beam curving described in Section \uppercase\expandafter{\romannumeral3}-A, the curved trajectory of the Airy beam is insufficient to circumvent such large obstacles, which consequently leads its performance to gradually converge toward that of the conventional focusing beam.

\begin{figure}[htbp]
    \centering
    \includegraphics[width=8.7cm]{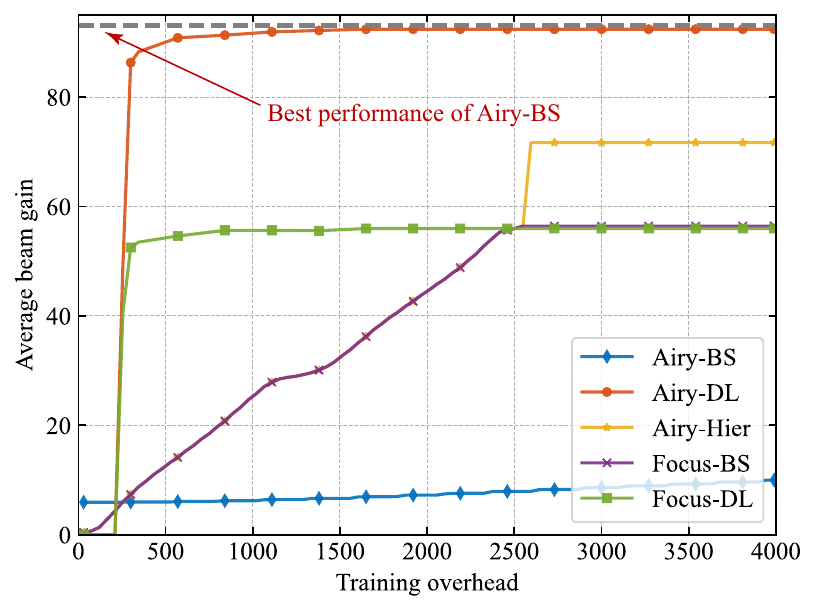}
    \caption{Average beam gain comparison with respect to the training overhead.}  
    \label{overhead}
\end{figure}
Fig. \ref{overhead} shows the average beam gain against the overhead. Since the training overhead of Airy-BS in our simulation setting reaches an exceptionally high value of 130,050, we present results only for training overhead ranging from 1 to 4,000 to ensure that all other methods, except Airy-BS, can achieve their optimal performance. The best performance of Airy-BS is indicated by the gray dashed line in the figure. Notably, an overhead of $n$ indicates that the optimal beam among the $n$ tested beams is selected for data transmission \cite{wch-twc, rainbow}, from which the corresponding beam gain is obtained. Although Airy-BS attains the highest optimal performance, its improvement with increasing training overhead is the slowest among all methods, primarily due to its exceptionally large codebook. This level of training overhead is clearly impractical in real-world applications. For Airy-Hier, which starts with a conventional angle and distance sweep, the initial performance trend resembles that of Focus-BS. Once the optimal angle and distance are identified, Airy-Hier performs an exhaustive search over the curvature at these parameters, resulting in better performance compared with the Focus-BS. However, because the angle and distance that yield the optimal Airy beam do not necessarily match those of the optimal focusing beam, its performance remains substantially lower than that of Airy-BS. In comparison, the proposed Airy-DL, with a training overhead of only 255, roughly $2\%$ of that required by Airy-BS, outperforms all other benchmarks and achieves performance close to the upper bound set by Airy-BS.

\begin{figure}[htbp]
    \centering
    \includegraphics[width=8.5cm]{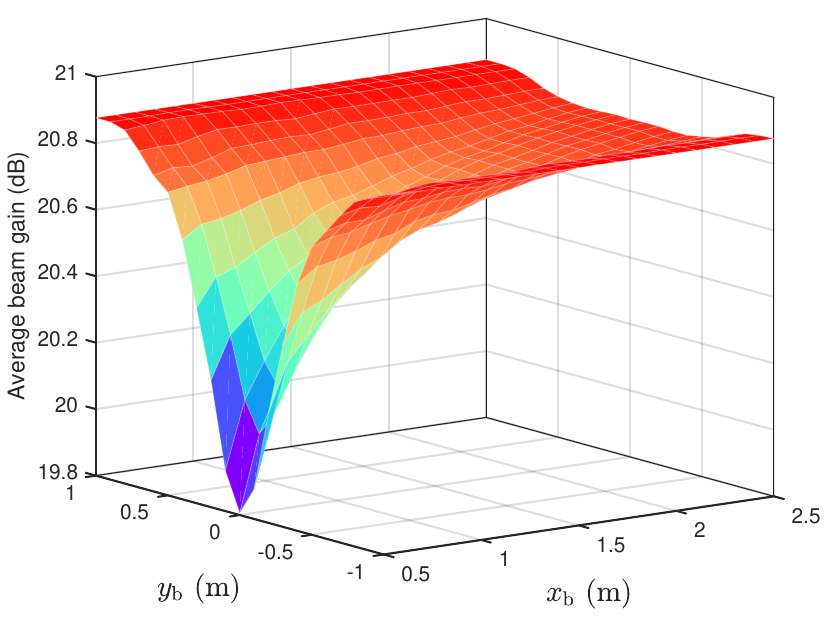}
    \caption{Average beam gain of the Airy-BS method versus obstacle center position.}  
    \label{location}
\end{figure}
Finally, Fig. \ref{location} takes the Airy-BS method as an example to illustrate the spatial distribution of the average beam gain when the obstacle is placed at different positions in the two-dimensional space. It is observed that the closer the obstacle is to the transmitter, the lower the average beam gain becomes. This is because obstacles located near the transmitter make the communication links in space more susceptible to blockage effects. Moreover, when the obstacle is positioned directly in front of the transmitter, the Airy beam is less capable of curving around it, which restricts its potential advantage. These findings provide practical insights for the placement of transmitter arrays in real-world deployments under fixed environmental layouts.

\section{Conclusion}
In this paper, we investigated an efficient and blockage-resilient near-field beam training scheme based on the self-accelerating Airy beam. We first derived the trajectory of the Airy beam under given parameters and obtained a closed-form expression in the paraxial regime. Then, we revealed an interesting finding that when a DFT codebook is employed for beam sweeping in the presence of obstacles, the received beam pattern not only reflects the receiver’s angle and distance but also implicitly encodes the relative angular and distance relationship between the receiver and the obstacle. Building upon this observation, the beam training task was formulated as a MTL problem, and a lightweight AMPBT-Net was proposed. This network takes a DFT beam pattern as input and jointly predicts the optimal Airy beam’s angle, distance, and curvature parameters. Simulation results demonstrate that, compared with conventional focusing beams, Airy beams effectively mitigate the detrimental impact of obstacles on the communication link between the transmitter and receiver. Moreover, the proposed deep learning–based beam training method substantially reduces the training overhead while maintaining performance comparable to exhaustive beam sweeping.

\begin{appendices}
\section{Proof of lemma 1}
Based on the beam profile in \eqref{eq4}, the electric field at point $(x,y)$ is rewritten as
\begin{equation}
\label{app_eq1}
E(x,y\,\vert \, E_0) =  \frac{1}{2\pi\sqrt{N}} \int_\mathcal{S} \frac{e^{j\Phi}x}{r_u^2}\left(j\kappa +\frac{1}{r_u}\right) dy_0,
\end{equation}
where $\Phi=\phi-\kappa r_u$ represents the total phase of the integrand. Compared with the total phase $\Phi$, the envelope of the integrand in \eqref{app_eq1}, i.e., $\frac{x}{r_u^2}\left(j\kappa+\frac{1}{r_u}\right)$ is slowly-varying. Hence, by applying the stationary-phase approximation \cite{stationary_phase_approximation}, which underlies the ray-optics interpretation of wave propagation, we can obtain the parametric expression of the trajectory in \eqref{eq15}. More specific, the condition of phase stationarity is that the first derivative of the total phase $\Phi$ with respect to $y_0$ is zero, i.e., $\Phi^{\prime}(y_0)=0$. This condition defines the ray emitted from point $y_0$ at the source aperture as 
\begin{equation}
\label{app_eq2}
\phi^{\prime}(y_0)+\kappa\frac{y_c-y_0}{r_c}=0,
\end{equation}
which contributes the main portion of the field observed at $(x,y)=(x_c,y_c)$ and $r_c=\sqrt{(y_c-y_0)^2+x_c^2}$. A caustic forms at the locus of points where adjacent rays converge. In the context of the stationary phase method, this corresponds to points where the phase function is not only stationary but also second-order stationary. This requires the second derivative of the phase function to be zero as
\begin{equation}
\label{app_eq3}
\phi^{\prime\prime}(y_0)-\kappa\frac{x_c^{2}}{r_c^3}=0.
\end{equation}

Equations \eqref{app_eq2} and \eqref{app_eq3} form a system that defines the caustic coordinates, which we denote $(x_c,y_c)$, parametrized by the source coordinate $y_0$. Explicitly, we first isolate the term $y_c-y_0$ from \eqref{app_eq2} as $y_c-y_0=-\frac{r\phi^{\prime}(y_0)}{\kappa}$, then we replace $y_c-y_0$ in $r_c=\sqrt{(y_c-y_0)^2+x_c^2}$, and we obtain the result
\begin{equation}
\label{app_eq4}
r_c^2=\frac{x_c^{2}}{\left(1-(\phi^{\prime}(y_0)/\kappa)^2\right)}.
\end{equation}
To obtain the expression of $x_c$ in \eqref{eq15}, we replace $r_c$ in \eqref{app_eq3} according to \eqref{app_eq4}, and we obtain the result
\begin{equation}
\label{app_eq5}
x_c=\frac{\kappa\left[1-(\phi^{\prime}(y_0)/\kappa)^2\right]^{3/2}}{\phi^{\prime\prime}(y_0)},
\end{equation}
and then the term $y_c$ is obtained by
\begin{equation}
\label{app_eq6}
\begin{aligned}
y_c&=y_0-\frac{r\phi^{\prime}(y_0)}{\kappa}=y_0-\frac{x_c\phi^{\prime}(y_0)}{\kappa\sqrt{\left(1-(\phi^{\prime}(y_0)/\kappa)^2\right)}}\\
&=y_0-\phi^{\prime}(y_0)\frac{1-(\phi^{\prime}(y_0)/\kappa)^2}{\phi^{\prime\prime}(y_0)}.
\end{aligned}
\end{equation}
Therefore, the proof of Lemma 1 is completed.

\section{Proof of lemma 2}
First of all, \eqref{eq5} can be rewritten as
\begin{equation}
\label{app_eq7}
    E(x,y\,\vert \, E_0) \overset{(a)}{\approx}\frac{1}{j\lambda} \int_\mathcal{S}  E_0(0,y_0) \frac{e^{-j\kappa r_u}x}{r_u^2}dy_0,
\end{equation}
where $(a)$ is because $\vert j\kappa \vert \gg \vert 1/r_u \vert$ under high-frequency bands. Next, by noting that $x=r_u\cos\theta^{\prime}$, where $\theta^{\prime}$ denotes the angle between the the propagation direction and the $y$-axis, \eqref{app_eq7} can be rewritten as
\begin{equation}
\label{app_eq8}
    E(x,y\,\vert \, E_0) \approx \frac{1}{j\lambda} \int_\mathcal{S}  E_0(0,y_0) \frac{e^{-j\kappa r_u}r_u\cos\theta^{\prime}}{r_u^2}dy_0.
\end{equation}
Under the paraxial assumption ($\theta^{\prime}\approx0$), we have $\cos\theta^{\prime}\approx1$, simplifying the expression to
\begin{equation}
\label{app_eq9}
    E(x,y\,\vert \, E_0) \approx \frac{1}{j\lambda} \int_\mathcal{S}  E_0(0,y_0) \frac{e^{-j\kappa r_u}}{r_u}dy_0.
\end{equation}
We then express the distance between the transmitting and receiving points as $r_u=\sqrt{x^2+(y-y_u)^2}$. For paraxial propagation where $x \gg \vert y - y_0 \vert$, the term $1/r_u$ can be approximated by $1/x$, leading to
\begin{equation}
\label{app_eq10}
    E(x,y\,\vert \, E_0) \approx \frac{1}{j\lambda x} \int_\mathcal{S}  E_0(0,y_0) e^{-j\kappa x\sqrt{1+\left(\frac{y-y_0}{x}\right)^2}}dy_0.
\end{equation}
Applying a first-order Taylor expansion to the square-root term in the exponent, we obtain
\begin{equation}
\label{app_eq11}
    E(x,y\,\vert \, E_0) \approx \frac{e^{-j\kappa x}}{j\lambda x} \int_\mathcal{S}  E_0(0,y_0) e^{-j\kappa \frac{\left(y-y_0\right)^2}{2x}}dy_0.
\end{equation}

Then, by substituting the beam profile in \eqref{eq4} into the electric field expression in \eqref{eq5}, and applying the stationary-phase approximation discussed in Appendix A, we can derive \eqref{eq16} as presented in Lemma 2. After performing some straightforward algebraic manipulations, the closed-form expression in \eqref{eq16} is obtained. For brevity, the detailed derivation steps are omitted here.

\end{appendices}

\bibliographystyle{IEEEtran}
\bibliography{Main}

\begin{thebibliography}{10}
\providecommand{\url}[1]{#1}
\csname url@samestyle\endcsname
\providecommand{\newblock}{\relax}
\providecommand{\bibinfo}[2]{#2}
\providecommand{\BIBentrySTDinterwordspacing}{\spaceskip=0pt\relax}
\providecommand{\BIBentryALTinterwordstretchfactor}{4}
\providecommand{\BIBentryALTinterwordspacing}{\spaceskip=\fontdimen2\font plus
\BIBentryALTinterwordstretchfactor\fontdimen3\font minus \fontdimen4\font\relax}
\providecommand{\BIBforeignlanguage}[2]{{%
\expandafter\ifx\csname l@#1\endcsname\relax
\typeout{** WARNING: IEEEtran.bst: No hyphenation pattern has been}%
\typeout{** loaded for the language `#1'. Using the pattern for}%
\typeout{** the default language instead.}%
\else
\language=\csname l@#1\endcsname
\fi
#2}}
\providecommand{\BIBdecl}{\relax}
\BIBdecl

\bibitem{THz-1}
W.~Jiang, Q.~Zhou, J.~He, M.~A. Habibi, S.~Melnyk, M.~El-Absi, B.~Han, M.~Di~Renzo, H.~D. Schotten, F.-L. Luo \emph{et~al.}, ``Terahertz communications and sensing for {6G} and beyond: A comprehensive review,'' \emph{IEEE Commun. Surv. Tuts.}, vol.~26, no.~4, pp. 2326--2381, 4th Quart. 2024.

\bibitem{THz-2}
M.~Li, J.~M. Jornet, D.~M. Mittleman, and C.~Han, ``Beam manipulation for terahertz communications: A new quality productive force,'' \emph{arXiv preprint arXiv:2503.22158}, 2025.

\bibitem{hop-1}
V.~Petrov, H.~Guerboukha, A.~Singh, and J.~M. Jornet, ``Wavefront hopping for physical layer security in {6G} and beyond near-field {THz} communications,'' \emph{IEEE Trans. Commun.,}, vol.~73, no.~5, pp. 2996--3012, May 2024.

\bibitem{THz-3}
Y.~Liu, Z.~Wang, J.~Xu, C.~Ouyang, X.~Mu, and R.~Schober, ``Near-field communications: A tutorial review,'' \emph{IEEE Open J. Commun. Soc.}, vol.~4, pp. 1999--2049, Aug. 2023.

\bibitem{THz-4}
I.~F. Akyildiz, C.~Han, Z.~Hu, S.~Nie, and J.~M. Jornet, ``Terahertz band communication: An old problem revisited and research directions for the next decade,'' \emph{IEEE Trans. Commun.}, vol.~70, no.~6, pp. 4250--4285, Jun. 2022.

\bibitem{bending_1}
S.~Droulias, G.~Stratidakis, and A.~Alexiou, ``Bending beams for {6G} near-field communications,'' \emph{IEEE Trans. Wireless Commun.}, vol.~24, no.~2, pp. 1467--1480, Feb. 2024.

\bibitem{sense_then_train}
H.~Jiang, Z.~Wang, and Y.~Liu, ``Sense-then-train: An active-sensing-based beam training design for near-field {MIMO} systems,'' \emph{IEEE Trans. Wireless Commun.}, vol.~23, no.~10, pp. 15\,525--15\,539, Oct. 2024.

\bibitem{wch-tvt}
C.~Weng, X.~Guo, and Y.~Wang, ``Near-field beam training with hierarchical codebook: Two-stage learning-based approach,'' \emph{IEEE Trans. Veh. Technol.}, vol.~73, no.~9, pp. 14\,003--14\,008, Sep. 2024.

\bibitem{wch-twc}
C.~Weng, X.~Guo, Y.~Guo, and Y.~Wang, ``Wavenumber domain beam training in {XL-MIMO} systems: Unifying far-field and near-field,'' \emph{IEEE Trans. Wireless Commun.}, 2025, early access.

\bibitem{HMIMO_1}
T.~Gong, L.~Wei, C.~Huang, G.~C. Alexandropoulos, M.~Debbah, and C.~Yuen, ``Near-field channel modeling for holographic {MIMO} communications,'' \emph{IEEE Wireless Commun.}, vol.~31, no.~3, pp. 108--116, Jun. 2024.

\bibitem{HMIMO_2}
Y.~Guo, X.~Guo, Y.~Chen, and Y.~Wang, ``Statistical channel estimation for holographic mimo exploiting the clustered sparsity,'' \emph{IEEE Trans. Veh. Technol.}, vol.~74, no.~8, pp. 13\,161--13\,166, Aug. 2025.

\bibitem{Rayleigh}
K.~T. Selvan and R.~Janaswamy, ``Fraunhofer and {Fresnel} distances: Unified derivation for aperture antennas,'' \emph{IEEE Antennas Propag. Mag.}, vol.~59, no.~4, pp. 12--15, Aug. 2017.

\bibitem{3gpp-122}
\BIBentryALTinterwordspacing
3GPP, ``Chair notes {RAN1}\#122 v14,'' {3rd Generation Partnership Project (3GPP)}, Tech. Rep., Sep. 2025. [Online]. Available: \url{https://www.3gpp.org/ftp/tsg_ran/WG1_RL1/TSGR1_122/Inbox/Chair_notes/Chair notes RAN1%23122 v14.zip}
\BIBentrySTDinterwordspacing

\bibitem{wavefront-1}
A.~Singh, V.~Petrov, P.~Sen, and J.~M. Jornet, ``Near-field terahertz communications for {6G} and beyond: From concepts to realizations,'' \emph{IEEE Signal Process. Mag.}, vol.~42, no.~1, pp. 106--125, Jan. 2025.

\bibitem{hop-2}
V.~Petrov, H.~Guerboukha, D.~M. Mittleman, and A.~Singh, ``Wavefront hopping: An enabler for reliable and secure near field terahertz communications in {6G} and beyond,'' \emph{IEEE Wireless Commun.}, vol.~31, no.~1, pp. 48--55, Feb. 2024.

\bibitem{Tcom-DLL}
M.~Cui and L.~Dai, ``Channel estimation for extremely large-scale {MIMO}: Far-field or near-field?'' \emph{IEEE Trans. Commun.}, vol.~70, no.~4, pp. 2663--2677, Apr. 2022.

\bibitem{hierarchical_1}
C.~Wu, C.~You, Y.~Liu, L.~Chen, and S.~Shi, ``Two-stage hierarchical beam training for near-field communications,'' \emph{IEEE Trans. Veh. Technol.}, vol.~73, no.~2, pp. 2032--2044, Feb. 2023.

\bibitem{learning-1}
W.~Liu, H.~Ren, C.~Pan, and J.~Wang, ``Deep learning based beam training for extremely large-scale massive {MIMO} in near-field domain,'' \emph{IEEE Commun. Lett.}, vol.~27, no.~1, pp. 170--174, Jan. 2022.

\bibitem{dft_bt_1}
X.~Wu, C.~You, J.~Li, and Y.~Zhang, ``Near-field beam training: Joint angle and range estimation with {DFT} codebook,'' \emph{IEEE Trans. Wireless Commun.}, vol.~23, no.~9, pp. 11\,890--11\,903, Sep. 2024.

\bibitem{dft_bt_2}
Z.~Wang, R.~Kiran, S.~Tsai, and R.~Zhang, ``Low-complexity near-field beam training with {DFT} codebook based on beam pattern analysis,'' \emph{arXiv preprint arXiv:2503.21954}, 2025.

\bibitem{gyq_training}
Y.~Guo, X.~Guo, and Y.~Wang, ``Cross {Rayleigh} and {Fresnel} distances: Unified far-field and near-field beam training for {XL-MIMO} using ellipse-fitting localization,'' \emph{IEEE Trans. Wireless Commun.}, 2025, early access.

\bibitem{GlobeCOM-NN}
H.~Chen, A.~Kludze, and Y.~Ghasempour, ``Curving around obstacles via {NN}-enabled wavefront shaping in {sub-THz} wireless networks,'' in \emph{Proc. of IEEE Globecom}, Cape Town, South Africa, Dec. 2024, pp. 5356--5362.

\bibitem{NC-princeton}
------, ``A physics-informed {Airy} beam learning framework for blockage avoidance in sub-terahertz wireless networks,'' \emph{Nature communications}, vol.~16, no.~1, p. 7387, 2025.

\bibitem{airy_codebook}
W.~Zhao, S.~Abadal, G.~Song, J.~Jiang, and C.~Han, ``Terahertz wireless data center: Gaussian beam or {Airy} beam?'' \emph{arXiv preprint arXiv:2504.20410}, 2025.

\bibitem{pinching_1}
Z.~Ding, R.~Schober, and H.~V. Poor, ``Flexible-antenna systems: A pinching-antenna perspective,'' \emph{IEEE Trans. Commun.}, 2025, early access.

\bibitem{pinching_2}
C.~Ouyang, Z.~Wang, Y.~Liu, and Z.~Ding, ``Array gain for pinching-antenna systems {(PASS)},'' \emph{IEEE Commun. Lett.}, vol.~29, no.~6, pp. 1471--1475, Jun. 2025.

\bibitem{movable_1}
L.~Zhu, W.~Ma, and R.~Zhang, ``Movable antennas for wireless communication: Opportunities and challenges,'' \emph{IEEE Commun. Mag.}, vol.~62, no.~6, pp. 114--120, Jun. 2023.

\bibitem{movable_2}
C.~Weng, Y.~Chen, L.~Zhu, and Y.~Wang, ``Learning-based joint beamforming and antenna movement design for movable antenna systems,'' \emph{IEEE Wireless Commun. Lett.}, vol.~13, no.~8, pp. 2120--2124, Aug. 2024.

\bibitem{RIS_1}
Y.~Liu, X.~Liu, X.~Mu, T.~Hou, J.~Xu, M.~Di~Renzo, and N.~Al-Dhahir, ``Reconfigurable intelligent surfaces: Principles and opportunities,'' \emph{Commun. Surv. Tuts.}, vol.~23, no.~3, pp. 1546--1577, 3rd Quart. 2021.

\bibitem{RIS_2}
B.~Zhao, Q.~Cui, W.~Ni, P.~Wang, Y.~Hou, and X.~Tao, ``Pareto optimization for performance trade-off in multi-functional {RIS}-assisted {ISAC} systems,'' \emph{IEEE Trans. Veh. Technol.}, 2025, early access.

\bibitem{RIS_3}
X.~Mu, J.~Xu, Y.~Liu, and L.~Hanzo, ``Reconfigurable intelligent surface-aided near-field communications for {6G}: Opportunities and challenges,'' \emph{IEEE Veh. Technol. Mag.}, vol.~19, no.~1, pp. 65--74, Mar. 2024.

\bibitem{commun_engineer}
H.~Guerboukha, B.~Zhao, Z.~Fang, E.~Knightly, and D.~M. Mittleman, ``Curving {THz} wireless data links around obstacles,'' \emph{Communications Engineering}, vol.~3, no.~1, p.~58, 2024.

\bibitem{cyb_wavenumber}
Y.~Chen, X.~Guo, G.~Zhou, S.~Jin, D.~W.~K. Ng, and Z.~Wang, ``Unified far-field and near-field in holographic {MIMO}: A wavenumber-domain perspective,'' \emph{IEEE Commun. Mag.}, vol.~63, no.~1, pp. 30--36, Jan. 2025.

\bibitem{airy_1}
N.~K. Efremidis, Z.~Chen, M.~Segev, and D.~N. Christodoulides, ``Airy beams and accelerating waves: an overview of recent advances,'' \emph{Optica}, vol.~6, no.~5, pp. 686--701, May 2019.

\bibitem{book_1}
M.~Born and E.~Wolf, \emph{Principles of optics: electromagnetic theory of propagation, interference and diffraction of light, 6th ed}, New York: Pergamon Press, 1980.

\bibitem{NirvaWave}
V.~Yazdnian and Y.~Ghasempour, ``Nirvawave: An accurate and efficient near field wave propagation simulator for {6G} and beyond,'' in \emph{Proc. IEEE Wireless Commun. Netw. Conf}, Milan, Italy, Mar. 2025, pp. 1--7.

\bibitem{beam_gain}
G.~Jiang and C.~Qi, ``Near-field beam training based on deep learning for extremely large-scale {MIMO},'' \emph{IEEE Commun. Lett.}, vol.~27, no.~8, pp. 2063--2067, Aug. 2023.

\bibitem{hierar_DLL}
Y.~Lu, Z.~Zhang, and L.~Dai, ``Hierarchical beam training for extremely large-scale {MIMO}: From far-field to near-field,'' \emph{IEEE Trans. Commun.}, vol.~72, no.~4, pp. 2247--2259, Apr. 2023.

\bibitem{stationary_phase_approximation}
I.~D. Chremmos, G.~Fikioris, and N.~K. Efremidis, ``Accelerating and abruptly-autofocusing beam waves in the {Fresnel} zone of antenna arrays,'' \emph{IEEE Trans. Antennas Propag.}, vol.~61, no.~10, pp. 5048--5056, Oct. 2013.

\bibitem{fast_bt}
Y.~Zhang, X.~Wu, and C.~You, ``Fast near-field beam training for extremely large-scale array,'' \emph{IEEE wireless Commun. Lett.}, vol.~11, no.~12, pp. 2625--2629, Dec. 2022.

\bibitem{THz-5}
C.~Han, Y.~Chen, L.~Yan, Z.~Chen, and L.~Dai, ``Cross far-and near-field wireless communications in terahertz ultra-large antenna array systems,'' \emph{IEEE Wireless Commun.}, vol.~31, no.~3, pp. 148--154, Jun. 2024.

\bibitem{MTAN}
S.~Liu, E.~Johns, and A.~J. Davison, ``End-to-end multi-task learning with attention,'' in \emph{Proc. IEEE/CVF Conf. Comput. Vis. Pattern Recognit}, 2019, pp. 1871--1880.

\bibitem{rainbow}
M.~Cui, L.~Dai, Z.~Wang, S.~Zhou, and N.~Ge, ``Near-field rainbow: Wideband beam training for {XL-MIMO},'' \emph{IEEE Trans. Wireless Commun.}, vol.~22, no.~6, pp. 3899--3912, Jun. 2022.

\end{thebibliography}

\vfill
\end{document}